\documentclass[sigconf]{acmart}

\renewcommand\footnotetextcopyrightpermission[1]{} 



\setcopyright{acmlicensed}
\acmPrice{15.00}
\acmDOI{10.1145/3540250.3549121}
\acmYear{2022}
\copyrightyear{2022}
\acmSubmissionID{fse22main-p428-p}
\acmISBN{978-1-4503-9413-0/22/11}
\acmConference[ESEC/FSE '22]{Proceedings of the 30th ACM Joint European Software Engineering Conference and Symposium on the Foundations of Software Engineering}{November 14--18, 2022}{Singapore, Singapore}
\acmBooktitle{Proceedings of the 30th ACM Joint European Software Engineering Conference and Symposium on the Foundations of Software Engineering (ESEC/FSE '22), November 14--18, 2022, Singapore, Singapore}

\usepackage[utf8]{inputenc}
\usepackage{xspace}
\usepackage{amsmath}
\usepackage{enumitem}
\usepackage{todonotes}

\usepackage{caption}
\usepackage{subcaption}

\allowdisplaybreaks
\usepackage{multirow}

\theoremstyle{definition}
\newtheorem{definition}{Definition}[section]

\usepackage{algorithm}
\usepackage{algpseudocode}
\algnewcommand\algorithmicmatch{\textbf{match}}
\algnewcommand\algorithmiccase{\textbf{case}}
\algdef{SE}[SWITCH]{Match}{EndMatch}[1]{\algorithmicmatch\ #1:}{\algorithmicend\ \algorithmicswitch}%
\algdef{SE}[CASE]{Case}{EndCase}[1]{\algorithmiccase\ #1}{\algorithmiccase}%
\algtext*{EndMatch}%
\algtext*{EndCase}%

\newcommand{\proto}{DeCon\xspace}

\usepackage{alltt}
\newenvironment{datalog}{\vspace{-0mm}\begin{alltt}\footnotesize}{\end{alltt}\vspace{-0mm}}

\colorlet{boxcolor}{teal!10!white}

\usepackage{listings}
\definecolor{verylightgray}{rgb}{.97,.97,.97}

\lstdefinelanguage{Solidity}{
	keywords=[1]{anonymous, assembly, assert, balance, break, call, callcode, case, catch, class, constant, continue, constructor, contract, debugger, default, delegatecall, delete, do, else, emit, event, experimental, export, external, false, finally, for, function, gas, if, implements, import, in, indexed, instanceof, interface, internal, is, length, library, log0, log1, log2, log3, log4, memory, modifier, new, payable, pragma, private, protected, public, pure, push, require, return, returns, revert, selfdestruct, send, solidity, storage, struct, suicide, super, switch, then, this, throw, transfer, true, try, typeof, using, value, view, while, with, addmod, ecrecover, keccak256, mulmod, ripemd160, sha256, sha3}, 
	keywordstyle=[1]\color{blue},
	keywords=[2]{address, bool, byte, bytes, bytes1, bytes2, bytes3, bytes4, bytes5, bytes6, bytes7, bytes8, bytes9, bytes10, bytes11, bytes12, bytes13, bytes14, bytes15, bytes16, bytes17, bytes18, bytes19, bytes20, bytes21, bytes22, bytes23, bytes24, bytes25, bytes26, bytes27, bytes28, bytes29, bytes30, bytes31, bytes32, enum, int, int8, int16, int24, int32, int40, int48, int56, int64, int72, int80, int88, int96, int104, int112, int120, int128, int136, int144, int152, int160, int168, int176, int184, int192, int200, int208, int216, int224, int232, int240, int248, int256, mapping, string, uint, uint8, uint16, uint24, uint32, uint40, uint48, uint56, uint64, uint72, uint80, uint88, uint96, uint104, uint112, uint120, uint128, uint136, uint144, uint152, uint160, uint168, uint176, uint184, uint192, uint200, uint208, uint216, uint224, uint232, uint240, uint248, uint256, var, void, ether, finney, szabo, wei, days, hours, minutes, seconds, weeks, years},	
	keywordstyle=[2]\color{teal},
	keywords=[3]{block, blockhash, coinbase, difficulty, gaslimit, number, timestamp, msg, data, gas, sender, sig, value, now, tx, gasprice, origin},	
	keywordstyle=[3]\color{violet},
	identifierstyle=\color{black},
	sensitive=false,
	comment=[l]{//},
	morecomment=[s]{/*}{*/},
	commentstyle=\color{gray}\ttfamily,
	stringstyle=\color{red}\ttfamily,
	morestring=[b]',
	morestring=[b]"
}

\lstdefinelanguage{DSC}{
    keywords=[1]{decl,public,violation},keywordstyle=[1]\color{violet},
	keywords=[2]{address, bool, byte, bytes, bytes1, bytes2, bytes3, bytes4, bytes5, bytes6, bytes7, bytes8, bytes9, bytes10, bytes11, bytes12, bytes13, bytes14, bytes15, bytes16, bytes17, bytes18, bytes19, bytes20, bytes21, bytes22, bytes23, bytes24, bytes25, bytes26, bytes27, bytes28, bytes29, bytes30, bytes31, bytes32, enum, int, int8, int16, int24, int32, int40, int48, int56, int64, int72, int80, int88, int96, int104, int112, int120, int128, int136, int144, int152, int160, int168, int176, int184, int192, int200, int208, int216, int224, int232, int240, int248, int256, mapping, string, uint, uint8, uint16, uint24, uint32, uint40, uint48, uint56, uint64, uint72, uint80, uint88, uint96, uint104, uint112, uint120, uint128, uint136, uint144, uint152, uint160, uint168, uint176, uint184, uint192, uint200, uint208, uint216, uint224, uint232, uint240, uint248, uint256, var, void, ether, finney, szabo, wei, days, hours, minutes, seconds, weeks, years},	
	keywordstyle=[2]\color{teal},
    keywords=[3]{r1,r2,r3,r4,r5,r6,r7,r8,r9,r10,r11,r12,r13,r14,r2'},
    keywordstyle=[3]\color{blue},
    keywords=[4]{sum},keywordstyle=[4]\color{teal},
	comment=[l]{//},
	morecomment=[s]{/*}{*/},
	commentstyle=\color{gray}\ttfamily,
}

\lstdefinelanguage{abstract}{
    keywords=[1]{insert,search,where,on},keywordstyle=[1]\color{violet},
	keywords=[2]{address, bool, byte, bytes, bytes1, bytes2, bytes3, bytes4, bytes5, bytes6, bytes7, bytes8, bytes9, bytes10, bytes11, bytes12, bytes13, bytes14, bytes15, bytes16, bytes17, bytes18, bytes19, bytes20, bytes21, bytes22, bytes23, bytes24, bytes25, bytes26, bytes27, bytes28, bytes29, bytes30, bytes31, bytes32, enum, int, int8, int16, int24, int32, int40, int48, int56, int64, int72, int80, int88, int96, int104, int112, int120, int128, int136, int144, int152, int160, int168, int176, int184, int192, int200, int208, int216, int224, int232, int240, int248, int256, mapping, string, uint, uint8, uint16, uint24, uint32, uint40, uint48, uint56, uint64, uint72, uint80, uint88, uint96, uint104, uint112, uint120, uint128, uint136, uint144, uint152, uint160, uint168, uint176, uint184, uint192, uint200, uint208, uint216, uint224, uint232, uint240, uint248, uint256, var, void, ether, finney, szabo, wei, days, hours, minutes, seconds, weeks, years},	
	keywordstyle=[2]\color{teal},
	comment=[l]{//},
	morecomment=[s]{/*}{*/},
	commentstyle=\color{gray}\ttfamily,
}

\newcommand{\new}[1]{{\color{black}#1}}

\titlenote{To appear at ESEC/FSE '22.}

\begin{document}

\title{Declarative Smart Contracts}

\author{Haoxian Chen}
\affiliation{%
  \institution{University of Pennsylvania}
  \country{USA}}
\email{hxchen@seas.upenn.edu}

\author{Gerald Whitters}
\affiliation{%
  \institution{University of Pennsylvania}
  \country{USA}}
\email{whitters@seas.upenn.edu}

\author{Mohammad Javad Amiri}
\affiliation{%
  \institution{University of Pennsylvania}
  \country{USA}}
\email{mjamiri@seas.upenn.edu}

\author{Yuepeng Wang}
\affiliation{%
  \institution{Simon Fraser University}
  \country{Canada}}
\email{yuepeng@sfu.ca}

\author{Boon Thau Loo}
\affiliation{%
  \institution{University of Pennsylvania}
  \country{USA}}
\email{boonloo@seas.upenn.edu}

\begin{abstract}

This paper presents {\em \proto}, a declarative programming language for
implementing smart contracts and specifying contract-level properties.
Driven by the observation that smart contract operations and contract-level
properties can be naturally expressed as relational constraints,
\proto models each smart contract as a set of relational tables that store transaction records.
This relational representation of smart contracts enables convenient specification
of contract properties, facilitates run-time monitoring of potential property violations,
and brings clarity to contract debugging via data provenance.
Specifically, a \proto program consists of a set of declarative rules and violation query rules
over the relational representation, describing the smart contract implementation and
contract-level properties, respectively.
We have developed a tool that can compile \proto programs into executable Solidity programs,
with instrumentation for run-time property monitoring.
Our case studies demonstrate that \proto can implement realistic smart contracts
such as ERC20 and ERC721 digital tokens.
Our evaluation results reveal the marginal overhead of \proto compared to the open-source reference implementation,
incurring $14\%$ median gas overhead for execution,
and another $16\%$ median gas overhead for run-time verification.

\end{abstract}

\maketitle

\section{Introduction}

Smart contracts are programs stored and executed on blockchains.
They have been used in a wide range of blockchain-enabled distributed 
applications to manage digital assets,
including auctions \cite{hahn2017smart}, financial contracts \cite{biryukov2017findel}, elections \cite{mccorry2017smart},
trading platforms \cite{notheisen2017trading}, and permission management \cite{azaria2016medrec}.
Unfortunately, today's smart contracts are error-prone and this has led to significant financial losses resulting from attacks such as
Dice2win \cite{Dice2win2018}, King of Ether \cite{king2016}, Parity Multisig Bug \cite{parity2017}, Accidental \cite{browne2017Accidental} and DAO \cite{dao2016,dao2016siegel}.

Over the past few years, different analysis and verification techniques have been proposed for known vulnerabilities of smart contracts, such as re-entrancy attacks and transaction-order dependency
\cite{grishchenko2018foundations,permenev2020verx,albert2018ethir,tsankov2018securify,schneidewind2020ethor}.
However, when it comes to high-level properties specific to individual
smart contracts, 
programmers typically have to 
rely on hand-written assertions~\cite{solidity-assert}, 
which is hard to maintain and error-prone.
For example, for a smart contract that manages digital 
tokens, one may want to ensure that 
all account balances add up to the total supply of tokens. 
To monitor this property during run-time,
one has to instrument the code to maintain a state that keeps track 
of the sum of all account balances, and add assertions about their equivalence
wherever either account balances or token supplies are updated.
\new{
There are third-party tools that support high-level property specification
and verification for Solidity, e.g., temporal logic~\cite{permenev2020verx}, 
formula with extended operators~\cite{li2020securing}, etc., 
but counter-examples are returned in the form of Ethereum bytecode traces 
or transaction sequences, which may not be easy for programmers to understand and 
localize bugs in the original implementation. 
}

To make smart contracts easier to analyze and verify,
we introduce {\em \proto}, a declarative programming language for
smart contract implementation and property specification.
\proto is based on Datalog~\cite{datalog}, a logic programming language.
Datalog frees programmers from low-level implementation details,
e.g., data structures, algorithms, etc.,
and allows them to reason about the contract on the specification level,
via inference rules \cite{bembenek2020formulog}.
In addition, such relational representation serves as a high-level abstraction 
of the contract, which enables efficient formal analysis and verification \cite{smaragdakis2010using,whaley2005using}.

A typical smart contract provides two kinds of interfaces: {\em transactions} 
and {\em views}. Transactions are function calls that alter the contract states,
e.g., a token transfer that updates both sender and recipient balances. 
Views are read-only functions that return
particular states of the contract, e.g., the balance of an account.

Smart contract properties and operations can be naturally mapped to 
relational logic.
For example, transactions, the main element in smart contracts, can be 
modeled as relational tables, where table schema contains transaction parameters, e.g.,
sender, recipient, and amount.
Similarly, the balance of each account can be expressed as sum aggregation on transaction records and
looking up an account balance can 
be expressed as a constraint on the address column of the balance table.

Given this relational view of transactions,
committing a transaction can be interpreted as appending a new row to
the corresponding table.
The commit and abortion logic of a pending transaction is specified
by declarative rules based on Datalog.
Views can be specified as declarative queries on these tables.
For example, an account balance is its total income subtracted by 
total expense, each of which is a query on relevant transaction records.

Contract properties are also specified as inference rules.
They are interpreted as property violation queries, a special kind of views, 
and are expected to be always empty during correct executions.
For example, if a smart contract forbids overspending, 
then a query on accounts with negative balances should always be empty.
Such unification of implementation and property specification language
saves programmers' effort to learn another language to formally specify properties.

\proto complies declarative specifications into executable Solidity \cite{solidity} programs
that run on blockchains, e.g., Ethereum,  and monitor the specified properties 
at run-time.
When a property (violation) view is derived non-empty 
after executing a pending transaction, the transaction is aborted.
Such automatic code generation not only saves implementation effort,
but also eliminates the gap between the program specification and 
implementation, providing a stronger guarantee of the verification result.

The key insight to generate efficient executable code from declarative 
specifications is that smart contract transactions are executed in sequence.
In other words, new rows are appended to the transaction tables one at a time.
Therefore, \proto borrows the idea of incremental view maintenance in 
databases~\cite{gupta1993maintaining} to generate efficient update procedures.
On committing a new transaction, instead of evaluating the queries
on the whole tables, only the differences in query results are computed 
and applied to existing views.

In addition, \proto is easy to debug with data provenance~\cite{buneman2001and}.
Provenance is a mechanism for explaining how certain tuples or 
facts are derived, right down to the input values. 
In an imperative language like Solidity~\cite{solidity},
dependency information is difficult to be captured automatically 
(through data-flow analysis). 
In contrast, inference rules in \proto give explicit dependency information,
where each tuple can be directly attributed to one rule,
thus providing more clarity to the execution process.

The key contributions of the paper are as follows:

\begin{itemize}
    \item We design \proto, a declarative language that unifies smart contract
    implementation and specification. We demonstrate its expressiveness 
    via case studies on representative smart contracts
    and their high-level correctness properties.
    \item We design an algorithm to compile these high-level 
    specifications into executable Solidity programs, with 
    instrumentation for run-time verification.
    \item We implement and experimentally evaluate \proto. 
    Our evaluation shows that the generated executable code 
    has comparable efficiency with the equivalent open-source 
    implementation of the same contract ($14\%$ median gas overhead),
    and the overhead of run-time verification is moderate
    ($16\%$ median gas overhead).
    The prototype implementation and evaluation benchmarks are
    open-sourced~\cite{dsc} for future studies and comparisons.
\end{itemize}

The rest of this paper is organized as follows.
Section~\ref{sec:example} motivates \proto using a Wallet example.
The declarative smart contract language is presented in Section~\ref{sec:language}.
Section~\ref{sec:compilation} demonstrates how to translate declarative rules into an executable Solidity program.
The expressiveness of \proto is demonstrated in Section~\ref{sec:case} using two case studies.
Section~\ref{sec:eval} experimentally evaluates \proto.
Section~\ref{sec:related} discusses related work, and
Section~\ref{sec:conc} concludes the paper.
\section{Illustrative example}
\label{sec:example}

In this section, we show how to use \proto to implement a smart contract,
specify its properties, and debug via provenance
using a Wallet smart contract that manages digital tokens.

\subsection{Contract Implementation}

A smart contract offers two kinds of interfaces: {\em transactions} and {\em views}.
Transactions are the function calls that update the contract states.
Views, on the other hand, are read-only functions that return one or more contract states.

In declarative smart contracts, transaction records are the only states.
Transactions are modeled as relational tables.
A new row is appended to the table when a new transaction is committed,
with column entries storing the transaction parameters.
Transaction rules, i.e., the condition on which a new transaction can be committed, are specified 
as declarative rules.
Finally, views are specified as declarative queries over the transaction tables.

We use the Wallet example, shown in Listing~\ref{lst:wallet}, to explain how 
relational tables and declarative rules can be specified.
The Wallet contract manages token transactions between Ethereum addresses,
where the contract owner can mint or burn tokens to addresses, and different 
addresses can transfer tokens to each other.

\begin{figure}[t]
\centering
\begin{lstlisting}[caption={Wallet smart contract},
label={lst:wallet}, frame=single,
numbers=left,
numbersep=5pt,
language=DSC
]
// Transaction event triggers 
.decl recv_mint(p:address, amount:int)
.decl recv_burn(p:address, amount:int)
.decl recv_transfer(from:address, to:address, amount:int)

// Views
.decl *totalSupply(n:int)
.decl balanceOf(p:address, n:int)[0]
.public totalSupply,balanceOf

// Transaction rules
.decl mint(p: address, amount: int)
.decl burn(p: address, amount: int)
.decl transfer(from: address, to: address, amount: int)
r1: mint(p,n):-recv_mint(p,n),msgSender(s),owner(s),n>0.
r2: burn(p,n):-recv_burn(p,n),msgSender(s),owner(s), 
               balanceOf(p,m), n<=m.
r3: transfer(s,r,n):-recv_transfer(s,r,n),balanceOf(s,m), 
                     n>=m.

// View rules
r4: totalSupply(n):-allMint(m),allBurn(b),n:=m-b.
r5: balanceOf(p,s):-totalOut(p,o),totalIn(p,i),s:=i-o.

// Auxiliary relations and rules ...
.decl totalMint(p: address, n: int)[0]
.decl totalBurn(p: address, n: int)[0]
r6: transfer(0,p,n) :- mint(p,n).
r7: transfer(p,0,n) :- burn(p,n).
r8: totalOut(p,s):-transfer(p,_,_),
                   s=sum n:transfer(p,_,n).
r9: totalIn(p,s):-transfer(_,p,_),
                  s=sum n:transfer(_,p,n).
.decl *allMint(n: int)
.decl *allBurn(n: int)
r10: allMint(s) :- s = sum n: mint(_,n).
r11: allBurn(s) :- s = sum n: burn(_,n).
\end{lstlisting}
\vspace{-15pt}
\end{figure}

\noindent \textbf{Relations and interfaces.}
Lines 1 to 14 declare the relations, with schema in the parenthesis,
and, optionally, primary key indices in the bracket 
(e.g., \lstinline{balanceOf} on line 8). Primary keys uniquely identify 
a row in the table. 
For instance, \lstinline{balanceOf} records the balance of each account,
and thus has unique account column.
Without explicit specification, all columns are treated 
as primary keys.
Relation \lstinline{totalSupply} (line 7) is a singleton relation,
a kind of relation that contains only one row and is
annotated by a star symbol.

Given these relation declarations, transaction and view interfaces are generated.
First, transaction interfaces are generated from relations with 
\lstinline{recv_} prefix,
where the the input parameters define the schema and
a boolean return value indicates the success of the transaction.
For example, relation \lstinline{recv_mint} is translated into the 
following interface in Solidity, the target executable language:
\begin{lstlisting}[frame=single,language=Solidity]
    function mint(address p, int amount) returns (bool);
\end{lstlisting}
Second, view functions are generated from the relations that appear 
in the public interface annotations (line 9). 
The input parameters are the primary keys, and output is the remaining 
values. Note that a singleton relation, e.g., \lstinline{totalSupply}, becomes
a function without parameters since it has no primary keys.
If all columns are primary keys, then the function returns a boolean 
value indicating the existence of the row.
For example, \lstinline{balanceOf(p:address, n:int)[0]} is translated into the following 
function interface:
\begin{lstlisting}[frame=single,language=Solidity]
    function balanceOf(address p) returns (int);
\end{lstlisting}

\noindent \textbf{Rules and functions.}
The rest of the program shows the rules that process transactions and
define the views.
Each rule is of the form \lstinline{<head> :- <body>}, 
interpreted as follows.
For all valuation of the variables that satisfies all constraints in the body, 
generate a row as specified in the head.
For example, \lstinline{r1} on line 15 says that a \lstinline{mint} transaction
can only be sent by the contract owner, and the amount should always greater
than $0$.
This rule is compiled into the following Solidity code
(with simplification):

\begin{lstlisting}[frame=single,language=Solidity]
function mint(address p, int n) (returns bool) {
    bool ret = false;
    if (msg.sender == owner && n>0) {
        // call functions to update dependent views...
        ret = true;
    }
    return ret;
}
\end{lstlisting}

When a \lstinline{mint} transaction is committed,
\lstinline{r5} will be triggered
through a chain of rules (\lstinline{r1->r6->r9->r5}).
It specifies the balance of an account \lstinline{p},
as the total income  \lstinline{totalIn(p,i)} subtracted by the total expense 
\lstinline{totalOut(p,o)},
with \lstinline{totalIn} and \lstinline{totalOut} further defined by \lstinline{r8}
and \lstinline{r9}, respectively.
This rule is compiled into two Solidity functions, each 
updates \lstinline{balanceOf[p]} when either \lstinline{totalIn} 
or \lstinline{totalOut} is updated.

\begin{lstlisting}[frame=single,language=Solidity]
function updateBalanceOfOnTotalIn(address p, int i) {
    int o = totalOut[p];
    balanceOf[p] = i-o;
}
function updateBalanceOfOnTotalOut(address p, int o) {
    int i = totalIn[p];
    balanceOf[p] = i-o;
}
\end{lstlisting}

To get the balance of a given account, 
one could call \lstinline{balanceOf}, a view function that takes 
the account address as parameter, and returns an integer value
as the account balance.
In \proto, relational tables are stored in maps, mapping primary
keys to values in remaining columns.
This view function is generated as follows.
\begin{lstlisting}[frame=single,language=Solidity]
function balanceOf(address p) public view returns (int) {
    // Read the row by primary key p
    BalanceOfTuple memory balanceOfTuple = balanceOf[p];
    // Return the value
    return balanceOfTuple.n;
}
\end{lstlisting}


\subsection{Specification and Run-time Verification}
\label{sec:example-monitor}

In \proto, properties are specified the same way as views,
but with additional annotation in order for \proto to know what to monitor at run-time.
For example, in the Wallet contract, one may want to make sure that
all account balances are always non-negative, which can be specified 
as follows.
\begin{lstlisting}[frame=single,language=DSC]
.decl negativeBalance(p:address,n:int)[0]
.violation negativeBalance
r14: negativeBalance(p,n) :- balanceOf(p,n), n < 0.
\end{lstlisting}

Rule \lstinline{r14} specifies the violation instance of the property:
for each row in \lstinline{balanceOf} table with \lstinline{n<0},
insert a row \lstinline{(p,n)} in \lstinline{negativeBalance} table.
During the execution of the transaction, the \lstinline{negativeBalance} table 
is incrementally updated when its dependent relations are updated, 
the same as other views.

A property is satisfied if and only if the {\em violation} table
is empty.
The keyword \lstinline{.violation} annotates that every
row in the \lstinline{negativeBalance} table is a property violation instance.
Given such annotations, \proto instruments the program to 
check the emptiness of all violation tables before each transaction is committed.

Note that properties are monitored on the granularity of transactions.
As we show in Section~\ref{sec:compilation}, due to the underlying
update procedure, transient violations could occur during the execution
of a transaction, but disappear at the end.
Therefore, instead of aborting right after a violation tuple is derived, 
a transaction is only aborted if, at the end of its execution, any violation 
table remains non-empty.
Such interpretation allows programmers to reason at the transaction level, 
without worrying about the underlying update procedure.

\new{
The violation checking procedure is generated and performed
at the end of each transaction. In this example, the 
\lstinline{negativeBalance} violation is checked as follows:
}
\smallskip
\begin{lstlisting}[frame=single,language=Solidity]
function checkViolations() {
    if negativeBalance is not empty:
        revert("Negative balance.")
    // check other violations ...
}
\end{lstlisting}

%
%

\subsection{Debugging via Provenance}

Data provenance is a feature of declarative programs that 
records the data flow from input to output and enables rule-wise debugging.
\new{It allows counter-example traces to be presented in the context of the original 
specification, instead of the low-level EVM instructions, 
thus making the debugging process more intuitive.
}

\begin{figure}
\centering
\includegraphics[width=0.27\textwidth]{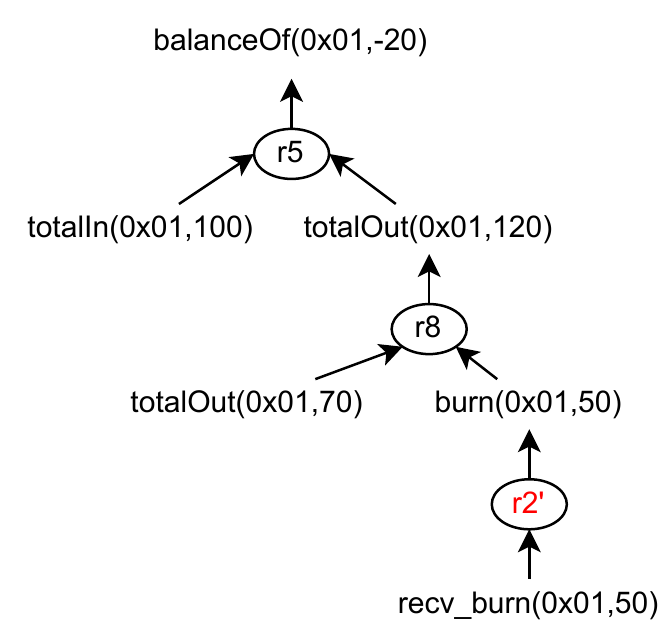}
\caption{Provenance of a violation of negative balance}
\label{fig:wallet-provenance}
\end{figure}

Suppose the original program has an incorrect $r2$, which
misses a predicate to check that the account has enough balance to be burnt.
The incorrect version of $r2$ is shown as $r2'$ in the following.
\begin{lstlisting}[frame=single,language=DSC]
r2': burn(p,n):-recv_burn(p,n),msgSender(s),owner(s). 
\end{lstlisting}
An account with balance $n$ would have negative balance if 
more than $n$ tokens are burnt.
Suppose during an execution, the account $0x01$ is detected to have 
negative balance of $-20$.
To understand why this violation happens, one could query the violation 
tuple's provenance tree, as shown in Figure~\ref{fig:wallet-provenance}.
The provenance tree is read from top to bottom.
On the top is a tuple \lstinline{balanceOf(0x01,-20)} that triggers 
the violation in \lstinline{negativeBalance}.
Below shows that it is derived by \lstinline{r5}, based on 
\lstinline{totalIn(0x01,100)} and \lstinline{totalOut(0x01,120)},
which are the total tokens received and sent by address \lstinline{0x01}.
The tuple \lstinline{totalOut(0x01,120)} is further derived by \lstinline{r8}.
This back tracing continues for another step until one finds
the derivation of \lstinline{r2'} is incorrect, which suggests that the 
condition \lstinline{balanceOf(p,m),m>=n} should be added.
With this provenance, programmers can debug contracts
in a visual and interactive manner.


\section{Language}
\label{sec:language}

A \proto contract consists of three elements:
relations, rules, and relation annotations.
A {\em relation} declaration specifies the name of a relational table and its schema.
Each relational table can store either transaction records, with the 
transaction parameters being the column values, or views,
the summary information of these transaction records.
A {\em rule} specifies either 
the conditions on which a new transaction gets approved
or the derivation of a view from the transaction records.
Finally, {\em relation annotations} specify whether a relational table is a public view 
or a violation.
Public views are compiled into public interfaces that take the relation's
primary keys as parameters, and return the remaining values in the matching row.
Violation will be monitored during run-time, and a transaction
is reverted if the violation relation is non-empty after the transaction execution.


\subsection{Relation Declarations and Annotations}

The formal syntax of relation declarations and annotations is defined
in Figure~\ref{fig:decl}.

\begin{figure}[!t]
\centering
\fcolorbox{black}{boxcolor}{
\begin{minipage}{0.46\textwidth}
\[
\begin{array}{rl}
    (Type)\; T        & :=\; int\;|~ uint\; |~ bool\; |~ address \\
    (Schema)\; S & :=\; c1:T1, c2:T2,... \\
    (Primary\;keys)\; K & :=\; k1,k2,... \\
    (Reserved\;relation)\; RS & \\
    (Singleton\;relation)\; SG & :=\; .decl\; *r(Schema) \\
    (Simple\;relation)\; SP & :=\; .decl\; r(Schema)[K] \\
    (Transaction\;relation)\; TR & :=\; .decl\; recv\_[r](Schema) \\
    (Relation)\; R  &  :=\; RS ~|~ SG ~|~ SP \\
    (Annotation)\; A & :=\; .public\; R \;|\; .violation\; R \\
\end{array}
\]
\end{minipage}}
\caption{Syntax of relation declarations and annotations}
\label{fig:decl}
\vspace{-4pt}
\end{figure}

\vspace{3pt}
\noindent \textbf{Schema.}
Schema of a relation is specified as a list of $c_i:T_i$,
where $c_i$ is the column name for the $i$th column,
and $T_i$ is the data type.

\vspace{3pt}
\noindent \textbf{Primary keys.}
Primary keys $K$ are a list of indices in the relation schema.
Primary key specification is optional. If a simple relation 
is specified without primary keys, then all columns are treated as primary keys.
Primary keys uniquely identify a row in each table.
On inserting a new row, if an existing row has the same primary key, 
the existing row is replaced by the new row.

\vspace{3pt}
\noindent \textbf{Singleton relations}
are relations with only one row, 
which are annotated with $*$ in the specification.
When a new row of a singleton relation is inserted, it replaces the existing row.

\vspace{3pt}
\noindent \textbf{Transaction relations}
are relations with prefix \lstinline{recv_}.
As explained in the next section,
when used in a rule, these relations are treated as an event trigger,
and are compiled into smart contract interfaces that handle incoming
transaction requests.

\vspace{3pt}
\noindent \textbf{Reserved relations.}
The following relations are reserved to handle smart contract specific constructs:
\begin{itemize}[leftmargin=*]
    \item \lstinline{msgSender(a:address)} stores the address of message sender.
    \item \lstinline{msgValue(v:uint)} stores the values of Ethers sent along a message.
    \item \lstinline{send(to:address, n:uint32)}
    triggers a transaction that sends $n$ \new{Ethers} to another account.
    \item \lstinline{constructor(*)} is translated into the constructor function,
    with schema being function parameters.
\end{itemize}

\subsection{Rules}

\begin{figure}[t]
\centering
\noindent \fcolorbox{black}{boxcolor}{\begin{minipage}{0.46\textwidth}
\begin{align*}
    (Variable)\; x & \\
    (Aggregation)\; Agg & := \; sum\;|\;max\;|\;min\;|\;count \\
    (Function)\; F & := \;  + | - | \times | \div \\
    (Condition)\; C & := \; > | < | >= | <= | == | != \\
    (Transaction\;relation)\; TR & :=\; recv\_[r] \\
    (Other\;relation)\; R & \\
    (Head\;literal)\; h & :=\; R(\bar{X}) \\
    (Body\;literal)\; b & :=\; R(\bar{X})\;|\; C(\bar{X}) \;|\; 
                                y = F(\bar{X})\\ 
                                &\quad\;|\; y = Agg\;x:R(\bar{X}) \\
    (Transaction\;Rule)\; Tx & :=\; h \;:-\; TR(\bar{X}), b_1, ..., b_n \\
    (View\;Rule)\; V & :=\; h \;:-\; b_1, ..., b_n \\
\end{align*}
\end{minipage}}
\caption{Syntax of rules}
\label{fig:rules}
\end{figure}

As shown in Figure~\ref{fig:rules}, we distinguish two kinds of rules: transaction rules and view rules.
A {\em transaction rule} contains a transaction relation in its body.
Transaction relations are relations with a prefix $recv\_$ in names.
These rules are only fired on receiving the corresponding transaction request,
and the transaction is approved if \new{the} rest of the constraints in the rule body
are satisfied.
A {\em view rule}, on the other hand, does not contain any transaction relations.
It is evaluated whenever one of the relations in the body is updated.

\noindent \textbf{Syntax restrictions.}
\proto does not support recursions. 
That is, no dependency loop exists between any two relations.
The dependency relationship in \proto is defined as follows.

\begin{definition}[Relation dependency]
\label{def:dependency}
Relation $R_a$ is dependent on relation $R_b$, if there exists a view 
rule where $R_a$ is in the head and $R_b$ is in the body,
or a transaction rule where $R_a$ is in the head, 
and $R_b$ is a transaction relation (with a prefix \lstinline{recv_}) 
in the body.
\end{definition}

\noindent \textbf{Rule semantics.}
A rule is evaluated as follows.
For each variables valuation $\pi$ that satisfies the {\em rule constraint},
generate the head tuple with all variables assigned to its corresponding values in $\pi$.
A variable valuation is a mapping from the set of variable names $V$ to the variable 
domain $D$ ($\pi: V \rightarrow D $).
Rule constraint is a conjunction of all body literal constraints.
As described in Figure~\ref{fig:rules}, there are four kinds of body literals.
For literals in the form of relational tuples $R(\bar{X})$, 
the constraint is satisfied if row $\bar{X}$ exists in the relational table $R$.
Other kinds of literals (i.e., conditions, functions, and aggregations) are directly 
interpreted as constraints on the variables.

Take \lstinline{r5} in the Wallet example (listing~\ref{lst:wallet}) for instance.
\begin{lstlisting}[frame=single,language=DSC]
r5: balanceOf(p,s):-totalOut(p,o),totalIn(p,i),s:=i-o.
\end{lstlisting}
This rule is interpreted as follows: "for all values of variable $p,o,i$ such that
there exists a tuple $totalOut(p,o)$ and $totalIn(p,i)$, derive the head tuple 
$balanceOf(p,s)$, where $s = i-o$".

\noindent \textbf{Aggregation literal}
$Agg\;x: R(\bar{X})$ computes the aggregate for all rows
in relation $R$ that satisfy the rule constraint.
For example, in the Wallet example (listing~\ref{lst:wallet}), 
line 31 shows a rule with aggregation.
\begin{lstlisting}[frame=single,language=DSC]
r8:totalOut(p,s):-transfer(p,_,_),s=sum n:transfer(p,_,n)
\end{lstlisting}
For each unique value of $p$ in the first column of \lstinline{transfer} table,
this rule computes the sum of the third column for rows in \lstinline{transfer} table
that has the value $p$ in the first column.
In other words, this rule groups the table by the first column,
and then computes the sum of the third column within each group.

\new{
\noindent \textbf{Limitations in expressiveness.}
\proto forbids recursions in order to keep the gas consumption predictable
and affordable. In fact, recursion is not recommended by the Solidity documentation 
for stack space issues~\cite{solidity-recursion}.

In addition, there are functions lie outside of relational logic, 
e.g. cryptographics, randomized functions, etc.
Such functions can be implemented in \proto by linking the contract
with external libraries. 
However, we recognize that analyzing such functions is a challenging problem,
and would need substantial future research.

Other unsupported features including contract inheritance and dynamic dispatching,
and checking interfaces of another contract, but they can be incorporated
into \proto in future compiler designs.
}

\section{Compilation to Solidity}
\label{sec:compilation}

\proto translates a set of declarative rules into an executable Solidity program
that
(1) processes transactions following the conditions in transaction rules, 
(2) updates views incrementally as new transactions are committed, and
(3) monitors property violations.

The compilation process involves three major steps. 

\noindent \textbf{(1) Abstract update functions.}
First, each rule is translated into a set of abstract update functions, 
each of which performs incremental updates to the head relation when 
one of the body relations is updated.
These functions are abstract in that they do not implement concrete 
data structures.
For example, recall that in the Wallet contract in Section~\ref{sec:example}, 
the following rule processes \lstinline{mint} transactions:
\begin{lstlisting}[frame=single,language=DSC]
r1: mint(p,n):-recv_mint(p,n),msgSender(s),owner(s),n>0.
\end{lstlisting}
It is translated into the following abstract update function:
\begin{lstlisting}[frame=single,numbers=left,numbersep=5pt,language=abstract]
    on insert recv_mint(p,n) {
      search owner where  {
        address s = owner.p;
        search msgSender where s==msg.sender {
          if(n>0) {
            insert mint(p,n)
          }}}}
\end{lstlisting}
This update function is triggered when a \lstinline{mint} transaction 
is received, as indicated by the event trigger tuple 
\lstinline{recv_mint(p,n)}.
The remaining two relational literals,
\lstinline{owner(s)} and \lstinline{msgSender(s)},
are translated into nested search statements (line 2 and line 4).
A search statement has the form \lstinline{(search R where C do S)}, 
where $R$ is the relational table, $C$ is the set of constrains on 
rows, and $S$ is the statement to execute for each row that satisfies
the constraints in $C$.
The condition literal (\lstinline{n>0}) is translated into an
\lstinline{if} statement (line 5).
If all prior conditions are satisfied, we arrive at line 6,
where the rule head is inserted.

\noindent \textbf{(2) Data structures.}
These abstract functions are then translated into concrete Solidity 
statements, where the \lstinline{search} statements become efficient
join algorithms on concrete data structures, 
and update functions for dependent views are called after 
an \lstinline{insert} statement.

\noindent \textbf{(3) Instrumentation.}
In the last step, the Solidity program is instrumented to monitor 
property violations, and abort the transaction if any violation
has been detected by the end of transaction execution.

\subsection{Abstract Update Function Generation}

There are two kinds of updates that could trigger a rule: 
tuple insertion and tuple deletion.
We use $Insert(e)$ and $Delete(e)$ to denote the update trigger
on inserting and deleting a tuple $e$, respectively.
Note that both a tuple and a literal have the form $R(\bar{X})$. 
It is called a tuple when $\bar{X}$ has concrete values,  
and called literal in a rule, where $\bar{X}$ is symbolic.
In the following discussion of update triggers we use literal and tuple
interchangeably.

Given a rule $r$, let $B(r)$ be the set of all relational literals 
in $r$'s body, and $e$ be the transaction relation in $r$ if $r$ is
a transaction rule, the set of update triggers $T(r)$ are defined as:
\begin{equation}
   T(r) \coloneqq 
   \begin{cases} 
        \bigcup\limits_{l\in B(r)} \{ Insert(l), Delete(l) \} 
        & r \text{ is View} \\
        \{Insert(e)\} & r \text{ is Tx rule} \\
   \end{cases}
\end{equation}
If $r$ is a view rule, then it can be triggered by updates of any
relation in its body. Otherwise, $r$ is a transaction rule, and 
it is only triggered on receiving a transaction request.

\begin{algorithm}
\caption{$\operatorname{UpdateFunction}(r,t)$. Given a rule $r$, and a trigger $t$,
returns an update object.}
\label{alg:update-function}
\begin{enumerate}
    \item Initialize the set of grounded variables $G \coloneqq t.variables$.
    \item Literals other than the trigger $L \coloneqq \{r.body \setminus t\}$.
    \item Update procedure $S \coloneqq \operatorname{Update}(r.head, L, t, G)$.
    \item Return (on t do S)
\end{enumerate}
\end{algorithm}

For each rule $r$, and for each update triggers in $T(r)$, 
an abstract update function is generated by 
$\operatorname{UpdateFunction(r,t)}$, presented in 
algorithm~\ref{alg:update-function}.
It first initializes the set of grounded variables by variables
in the trigger literal.
Grounded variables are variables that are constrained to a 
constant value.
Variables in a trigger literal are considered grounded 
because the update function is always triggered by the insertion 
or deletion of a concrete tuple.
In step(3), update procedure $S$ is generated by a sub-routine
$\operatorname{Update}$, which is presented in algorithm~\ref{alg:update}.
Finally, it returns the abstract update function in the form of 
(\lstinline{on t do S}), where $t$ is the update trigger and 
$S$ is the update procedure.

\begin{algorithm}
\caption{$\operatorname{Update}(h,L,t,G)$. Given a rule head $h$, 
a list of body literals $L$, an update trigger $t$, 
and the set of grounded variables $G$,
return statements that perform the incremental update.}
\label{alg:update}
\begin{algorithmic}
\Match{$L$}
    \Case{$Nil$ => \textbf{match} t} 
            \Case{Insert => \Return Insert(h)}
            \EndCase
            \Case{Delete => \Return Delete(h)}
            \EndCase
    \EndCase
    \Case{$head::tail$ =>}
        \State Add grounded variables 
            $G' \coloneqq G \cup \{x|x \in head \} $
        \State Inner statements
            $S \coloneqq \operatorname{Update(h,tail,t,G')}$
        \Match{head} 
            \Case{$R(\bar{X})$ =>}
                \State Derive constraints  
                    $C \coloneqq \operatorname{Constraint}(R(\bar{X}),G) $
                \State \Return (Search R where C do S)
            \EndCase
            \Case{$C(\bar{X})$ => \Return (If C Then S) }
            \EndCase
            \Case{$y=F(\bar{X})$ => \Return ($y=F(\bar{X})::S$)}
            \EndCase
            \Case{$y=Agg\;x:R(\bar{X})$ =>} 
                \State \Return ($y=Agg\;x:R(\bar{X})::S$)
            \EndCase
            
        \EndMatch
    \EndCase
\EndMatch 
\end{algorithmic}
\end{algorithm}

As shown in Algorithm~\ref{alg:update}, $\operatorname{Update(h,L,t,G)}$
performs recursion on $L$, the list of literals in the rule body,
with every recursion translating one literal to a layer of code block,
nested within the code block generated by the previous literals.

In particular, it performs pattern matching on input $L$, 
a list of literals to be translated.
If $L$ is empty, which means all body literals have been translated,
an update statement consistent with the update trigger is returned.
Otherwise, $L$ has the form $head::tail$.
It first adds all variables in $head$ into the set of grounded variables,
and then generates the inner code blocks $S$ by recursively calling
itself on $tail$ and the updated set of grounded variables $G'$.
Depending on the form of $head$, the current layer of code block 
is generated in different ways.
By the syntax of the language in Section~\ref{sec:language}, 
$head$ could take one of the following forms:
\begin{itemize}[leftmargin=*]
    \item A relational literal $R(\bar{X})$.
        Given the set of grounded variables $G$, 
        the search constraints for rows in $R$ 
        is generated as follows.
        \begin{equation*}
            \begin{aligned}
            Constraints(R(\bar{X}),G) \coloneqq \bigwedge 
                    & \{ (R[i]==v) | v \in G, v \in \bar{X}, \\
                    & i = \bar{X}.\operatorname{indexOf}(v) \}
            \end{aligned}
        \end{equation*}
        where $R[i]==v$ means filtering rows in table $R$ whose 
        $i$-th column equals to $v$.
    \item A condition literal $C(\bar{X})$, in which case,
        the condition is directly used in the same way as an {\sf If} condition,
        with the inner code block $S$ placed within the {\sf If}
        statement.
    \item A function or aggregation. In either case, the
        literal is directly translated into an assignment
        statement, followed by the inner code block $S$.
\end{itemize}

\noindent \textbf{Aggregations.}
The evaluation results of aggregation functions are maintained
incrementally.
Sums are incremented by $n$ when a row with aggregate 
value $n$ is inserted, and decremented by $n$ when a row is deleted.
Similarly, counts are incremented by $1$ on row insertion,
and decremented by $1$ on row deletion.
Maximums and minimums are slightly different.
When a new row is inserted with value $n$, if $n$ is greater than the 
current maximum, the maximum is updated to $n$.
When the current maximum row is deleted, the maximum is updated
as the second maximum value. 
Thus, it requires maintaining a sorted list of values.
Minimum is maintained in a similar fashion.

\subsection{Concrete Data Structures and Instructions}
\label{sec:data-structure}

Given the abstract functions generated from each rule,
the next step is to generate concrete and efficient data
structures and search algorithms in the Solidity language.

\vspace{3pt}
\noindent \textbf{Data structures.}
Each relational table $R$, except singleton relations, 
is translated into a mapping from its primary keys 
to a structure that stores the rest of the column values:
\begin{lstlisting}[frame=single,language=Solidity]
    struct RTuple {
        bool valid;
        T1 field1;
        T2 field2;
        ...
    };
    mapping(k1 => k2 => ... => kn => RTuple) R;
\end{lstlisting}
Note that hash-maps in Solidity by default map all keys to zero.
Therefore, a valid bit (\lstinline{valid}) is introduced to indicate 
the existence of a tuple.
Columns other than primary keys are the structure members.
If all columns are primary keys, then its structure only contains a valid bit.

Singleton relations are directly stored in a structure with columns
being the structure members.

\vspace{3pt}
\noindent \textbf{Join index.}
Join index is built for each search statement in the abstract 
update program.
Given a search statement \lstinline{Search R where C do S} 
in the abstract update program, 
if all primary keys of $R$ are constrained to constant values,
no join index is generated.
The matching entry can be directly looked up by primary keys.

On the other hand, if, in some rules, 
not all primary keys of $R$ are constrained to constant values, 
a join index is built as a map from the constrained keys to 
a list of unconstrained keys.

Suppose relation $R_1(k1,k2,v1)$ has two primary keys $k1$ and $k2$. 
As described above, table $R1$ is stored as a map from primary keys
to remaining values (\lstinline{mapping(k1=>k2=>R1Tuple)}).
Given a search statement \lstinline{Search R1 where R1[0]==k1 do S},
where only one primary key $k1$ is constrained,
the join index for $R_1$ is built as the following.

\begin{lstlisting}[frame=single,language=Solidity]
    struct R1KeyTuple {
        bool valid;
        T2 k2;
    }
    mapping(k1 => R1KeyTuple[]) R1Index;
\end{lstlisting}

where \lstinline{R1Index} maps $k1$ to a list of $R1KeyTuple$,
which stores value of the other primary key $k2$. 
During the join execution,
to iterate all rows in $R_1$ that satisfy $R_1[0]==k1$,
it first looks up all $k2$ in \lstinline{R1KeyTuple[k1]},
and then for each $k2$, get the value in \lstinline{R1[k1][k2]}.

\medskip
\noindent \textbf{Update dependent views.}
An insert or delete statement in the abstract update function
is translated into two sets of Solidity instructions:
(1) update the corresponding data structure; and (2) call the update functions
for the dependent relations (Definition~\ref{def:dependency}).

Inserting a relational tuple $t_1$ directly updates the map,
as well as the join index if one exists.
If a tuple $t_0$ with the same primary keys exists, 
all dependent views are updated by first calling deletion updates on $t_0$, 
and then the insertion updates on $t_1$.
Insertion update refers to functions triggered by tuple insertion, 
and deletion update refers to functions triggered by tuple deletion.
Otherwise, insertion updates are directly called.
Since a Solidity mapping maps all keys to value zero by default, 
a tuple exists if its valid bit is set to true.

Deleting a relational tuple resets its valid bit to false.
Then deletion updates are called for all dependent relations.

In this way, when a new transaction is committed, 
all dependent views are updated through 
this chain of update propagation.
Since there is no recursion, i.e., dependency loop between relations,
allowed in \proto, update propagation is guaranteed to terminate.

\medskip
\noindent \textbf{Logging.}
\new{
Committed transactions are logged as Solidity Events~\cite{solidity-events},
a more gas efficient storage than global memory, but can only be read offline.
}
These events constitute all states of a \proto contract,
which enable offline analysis for further insights and potential bugs.


\subsection{Run-time Verification}
\label{sec:instrument}

Properties are specified as declarative rules that derive
violation instances.
Such relations are annotated with the keyword \lstinline{violation}.


\new{
As introduced in Section~\ref{sec:example-monitor}, transient violations 
that occur during the transaction execution are not counted.
To see why transient violations can occur,
}
consider again the Wallet contract in 
Section~\ref{sec:example}, and a property that all account balances
add up to the total supply. The property can be specified as shown in Listing~\ref{lst:balance}.

\bigskip
\begin{lstlisting}[frame=single,
language=DSC,
label={lst:balance},
caption={All account balances add up to total supply.}]
.violation unequalTotalSupply
r12: totalBalance(s) :- sum n: balanceOf(_,n).
r13: unequalTotalSupply(n,m):-totalSupply(n),
                              totalBalance(m),n!=m.
\end{lstlisting}

During the execution of a \lstinline{mint} transaction, 
the \lstinline{totalSupply} and the \lstinline{totalBalance}
are updated in sequence, which leads to a violation 
when one is updated before another, but the violation disappears when both are
updated.

Given this notion of transient violations,
instead of aborting the transaction right after a violation tuple
is derived, the checking procedure is deferred to
the end of transaction.
If any violation view is non-empty,
the transaction is aborted.
Note that a Solidity mapping does not record its \new{domain}.
So a separate array of mapping keys are maintained
and iterated for valid violation tuples.

\subsection{Provenance Generation}

To debug a violation, programmers can use data provenance to visualize 
the derivation process of a violation tuple.
As shown in Figure~\ref{fig:wallet-provenance}, provenance is a
directed graph with two kinds of vertices: tuples and rules.
Edges from a tuple vertex to a rule vertex denote tuple reads,
and edges from a rule vertex to a tuple vertex denote tuple
derivations.

To generate this provenance graph during the rule evaluation procedure,
two kinds of additional records are logged:
tuple read \lstinline{Read(tuple, rid)} and tuple derivation 
\lstinline{Write(rid,tuple)}, where \lstinline{rid} is a unique
identifier for each rule.
\lstinline{Read(tuple,rid)} is interpreted as an edge
from \lstinline{tuple} to the rule indexed by \lstinline{rid},
and, conversely, \lstinline{Write(rid,tuple)} is an edge from
rule \lstinline{rid} to \lstinline{tuple}.

Note that in Solidity, a failed transaction reverts all instructions, 
including logging.
When a transaction is reverted due to a property violation,
the provenance logs would also be reverted.
Therefore, to generate provenance for a violation tuple, 
the transaction needs to be executed in a local debugging environment
instead of the deployment blockchain.
This practice also saves storage space on the public blockchain.
\subsection{Optimizations}
\label{sec:optimizations}
To improve gas and storage efficiency, two optimizations
have been applied to the generated codes.

\medskip
\noindent \textbf{Join order.}
Body literals in a rule
are sorted by their iteration cost in an increasing order.
First are reserved relations and singleton relations, since they  
need no iteration.
Second are the relations whose primary keys have all appeared in 
proceeding literals.
These literals can be searched via a direct mapping look-up, thus
requiring no iterations either.
Next are the rest of the relations, which are translated into loops.
Finally come condition and function literals.

\medskip
\noindent \textbf{Storage space.}
Storage space on a blockchain is precious due to the high synchronization cost.
Deriving relations on-demand, that is, delaying evaluating an inference rule 
until it is used, can save storage space, but may incur performance overhead.
To achieve a balanced trade-off between time and space,
\proto only proactively derives and stores relations annotated as public views or 
violations, as well as relations that are read during their derivation.
Other relations are derived on-demand.
For example, in the Wallet contract in Section~\ref{sec:example},
relation \lstinline{mint} only serves as an update trigger 
for dependent rules, which is never queried during the update of public 
views or violations. Therefore, when a \lstinline{mint} tuple is generated by
\lstinline{r1}, it only triggers the update for dependent rules,
but it is not written to the persistent storage.
\section{Case studies}\label{sec:case}

In this section we demonstrate the expressiveness of \proto
and the explainability of data provenance via case studies on
two typical smart contracts.
For brevity, only a subset of rules are discussed.
\new{
All contracts are available online~\cite{dsc-benchmarks}.
}

%

\subsection{ERC20}

ERC20~\cite{erc20} is a token standard for fungible tokens.
Similar to the Wallet contract in Section~\ref{sec:example},
it also supports token transfers between users.
In addition, it has an allowance mechanism, where users
can allow other users to transfer their tokens, 
\new{
up to a certain amount called \lstinline{allowance}.
}
The allowance mechanism can be specified as follows.
\begin{lstlisting}[frame=single,language=DSC]
r1: transferFrom(sender,receiver,spender,n) :- 
        recv transferFrom(sender, receiver, n),
        /* Sender has enough balance. */
        balanceOf(sender,m), m>=n, 
        /* Operator has enough allowance. */
        msgSender(spender),
        allowance(sender,spender,l),l>=n.
\end{lstlisting}

On receiving a \lstinline{transferFrom} transaction,
in addition to checking that the \lstinline{sender} has enough balance 
(\lstinline{m>=n}), the rule also requires
the \lstinline{spender} to have enough allowance to spend
tokens on \lstinline{sender}'s behalf (\lstinline{l>=n}).
The relation \lstinline{transferFrom} 
represents transactions where the \lstinline{spender} sends $n$ 
tokens from the \lstinline{sender} to the \lstinline{receiver}.

The allowance of a \lstinline{spender} on an account is
specified as follows.
\begin{lstlisting}[frame=single,language=DSC]
r2: spentTotal(o,s,m) :- transferFrom(o,_,s,_), 
                    m = sum n: transferFrom(o,_,s,n).
r3: allowance(o,s,n) :- allowanceTotal(o,s,m), 
                        spentTotal(o,s,l), n := m-l.
\end{lstlisting}
The relation \lstinline{spentTotal} accounts
the amount of tokens \lstinline{m} that spender \lstinline{s} has spent on
behalf of the sender \lstinline{o}.
And \lstinline{allowance} is derived by subtracting
the total spending from the total allowance, 
an amount approved by the sender (defined in another rule).

Given the definition of \lstinline{allowance} and the \lstinline{spentTotal},
we can specify a property that a spender never overspends as the following:
\begin{lstlisting}[frame=single,language=DSC]
.violation overSpent
overSpent(o,s,n,m) :- allowanceTotal(o,s,n),
                      spentTotal(o,s,m), m>n.
\end{lstlisting}
\proto then generates instrumentation to monitor this property at run-time.


\noindent \textbf{Explain allowance changes via data provenance.}
Suppose the programmer made a mistake in specifying \lstinline{spentTotal}: 
\begin{lstlisting}[frame=single,language=DSC]
r2': spentTotal(o,s,m):-transferFrom(o,_,s,_), 
                   m = sum n: transferFrom(_,_,s,n). 
\end{lstlisting}
The error is in the \lstinline{sum} literal, where \lstinline{transferFrom} records
should be grouped by both \new{sender} \lstinline{o} and spender \lstinline{s},
instead of just the spender.

When a spender account $s$ wants to transfer 20 tokens from account $a$ to
$b$, by submitting a transaction $\operatorname{transferFrom(a,b,s,20)}$,
it is reverted.
\proto explains that it is because the condition $l >= n$ in $r1$ is false,
which means the spender $s$ does not have sufficient allowance to transfer
tokens on $a$'s behalf.

\begin{figure}
    \centering
    \begin{subfigure}[t]{0.23\textwidth}
        \centering
        \includegraphics[width=\textwidth]{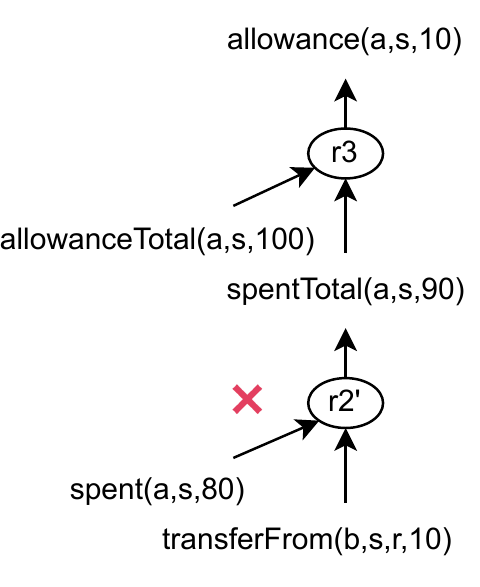}
        \caption{allowance(a,s,10)}
        \label{fig:provenance}
    \end{subfigure}%
    ~
    \begin{subfigure}[t]{0.23\textwidth}
        \centering
        \includegraphics[width=\textwidth]{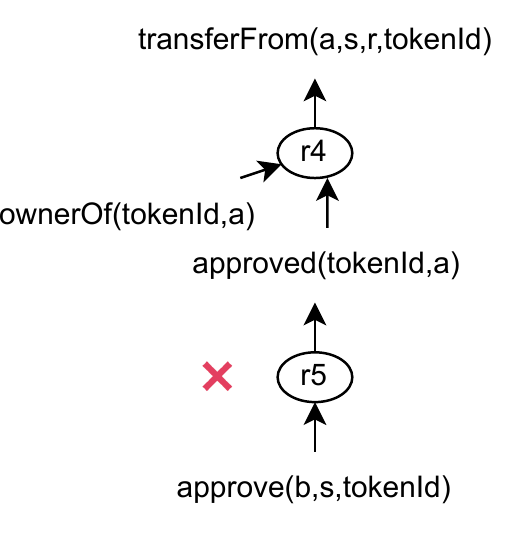}
        \caption{transferFrom(a,s,r,tokenId)}
        \label{fig:provenance-721}
    \end{subfigure}
    \caption{Provenance tree for tuples.}
    \label{fig:my_label}
\end{figure}

To understand why the \lstinline{spender} only have 10 allowance to $a$'s account,
one could get the provenance of the tuple \lstinline{allowance(a,s,10)}, 
as shown in Figure~\ref{fig:provenance}. 
On top of the provenance tree, \lstinline{allowance(a,s,10)} is derived by 
\lstinline{r3},
from the fact that the total allowance is 100 (\lstinline{allowanceTotal(a,s,100)}), 
and that \lstinline{a} has spent $90$ already (\lstinline{spentTotal(a,s,90)}).
To see why \lstinline{spentTotal(a,s,90)} is derived, 
the programmer continues expanding its provenance tree.
A bug is revealed at this step, where
a transaction from address $b$ to $r$ is accounted
for $s$'s allowance on address $a$, which points to the bug in $r2'$.

\subsection{ERC721}

ERC721~\cite{erc721} is a smart contract standard for non-fungible tokens (NFTs).
A main transaction for ERC721 tokens is \lstinline{transfer}, which
records the transfer of a token from \lstinline{sender} to 
\lstinline{recipient} at a particular \lstinline{time}.
The transaction time is included to specify the following views.

First is the view function \lstinline{ownerOf}.
Given the \lstinline{transfer} relation,
the owner of a token is defined as follows.
\begin{lstlisting}[frame=single,language=DSC]
latestTransfer(tokenId,s,r,t) :- transfer(tokenId,s,r,t),
                      t = max i: transfer(tokenId,_,_,i).
ownerOf(tokenId, p):-latestTransfer(tokenId,_,p,_),p!=0.
\end{lstlisting}
where the first rule selects the latest transfer record for 
\lstinline{tokenId},
and the next rule specifies that if the recipient of the 
latest transfer is non-zero, it is the owner of the token.

Next is the \lstinline{exist} relation.
A token exists if it is minted and is not burnt.
In ERC721 contracts, 
burning a token emits a transfer record from its owner to zero address.
So \lstinline{exist} is defined as:
\begin{lstlisting}[frame=single,language=DSC]
exists(tokenId, true) :- 
    latestTransfer(tokenId,_,to,_), to!=0.
\end{lstlisting}
The rule checks that a token's latest transfer recipient is a non-zero address, 
which means it is not burnt.

To ensure every existing token has an owner, we could specify the following
property:
\begin{lstlisting}[frame=single,language=DSC]
.violation tokenNoOwner
tokenNoOwner(tokenId) :- ownerOf(tokenId,o), o==0.
\end{lstlisting}
which defines a property violation as entries in the \lstinline{ownerOf}
table with owner address \lstinline{0}.

\noindent \textbf{Explain an unexpected token transfer via data
provenance.}
Suppose the owner wants to understand why one of her tokens
has been transferred away in a transaction 
\lstinline{transferFrom(a,s,r,tokenId)}, 
\new{
where \lstinline{a} is the operator,
\lstinline{s} is the sender, and \lstinline{r} is the receiver,
}
she expands the 
provenance tree for the transaction, 
which is shown in Figure~\ref{fig:provenance-721}.
On top of the provenance tree is a \lstinline{transferFrom} tranaction,
approved by the following rule:
\begin{lstlisting}[frame=single,language=DSC]
r4: transferFrom(operator, sender, receiver, tokenId) :- 
        recv transferFrom(operator, sender, receiver, tokenId),
        /* Sender owns the token. */
        ownerOf(tokenId, sender), 
        /* Operator is approved to move the token. */
        msgSender(operator), approved(tokenId,operator).                        
\end{lstlisting}
where $approved(tokenId,operator)$ means that the token
\lstinline{tokenId} has been approved to use by \lstinline{operator}.
This approval is set by the token owner.

Suspicious about the \lstinline{approved(tokenId,a)} tuple,
the owner continues to expand the provenance tree,
and finds that it is derived from the following rule:
\begin{lstlisting}[frame=single,language=DSC]
r5: approved(tokenId,operator) :-
                    approve(_,operator,tokenId).
\end{lstlisting}
and the tuple \lstinline{approve(b,s,tokenId)}, which means
account \lstinline{b}, a previous owner, has approved this token
to operator \lstinline{s} before transferring this token to \lstinline{a}.
Here, she finds the bug: \lstinline{r5} does not
check that the address that approves the token should be the token owner.
The rule should have been updated as follows instead:

\begin{lstlisting}[frame=single,language=DSC]
r5': approved(tokenId,operator):-ownerOf(tokenId,owner), 
        approve(owner,operator,tokenId).
\end{lstlisting}

\begin{table*}[]
\begin{tabular}{l|lll|ll|l|lll}
\hline
\multirow{2}{*}{Contract}      & \multirow{2}{*}{LOC} & \multirow{2}{*}{\# Functions} & \multirow{2}{*}{\# Rules} & \multicolumn{2}{|l|}{Byte-code size (KB)}   & \multirow{2}{*}{Transaction} & \multicolumn{2}{c}{Gas cost (K)} &       \\
                               &                      &                               &                           & Reference           & \proto                 &                              & Reference       & Compiled       & Diff  \\
\hline
\multirow{3}{*}{Wallet}        & \multirow{3}{*}{57}  & \multirow{3}{*}{6}            & \multirow{3}{*}{12}       & \multirow{3}{*}{3}  & \multirow{3}{*}{3}  & mint                         & 36              & 62             & 70\%  \\
                               &                      &                               &                           &                     &                     & burn                         & 36              & 47             & 29\%  \\
                               &                      &                               &                           &                     &                     & transfer                     & 52              & 38             & -26\% \\
\hline
\multirow{4}{*}{Crowdsale}     & \multirow{4}{*}{70}  & \multirow{4}{*}{5}            & \multirow{4}{*}{11}       & \multirow{4}{*}{4}  & \multirow{4}{*}{3}  & invest                       & 38              & 33             & -12\% \\
                               &                      &                               &                           &                     &                     & close                        & 38              & 47             & 25\%  \\
                               &                      &                               &                           &                     &                     & withdraw                     & 26              & 29             & 14\%  \\
                               &                      &                               &                           &                     &                     & claimRefund                  & 29              & 33             & 13\%  \\
\hline
\multirow{3}{*}{SimpleAuction} & \multirow{3}{*}{139} & \multirow{3}{*}{3}            & \multirow{3}{*}{13}       & \multirow{3}{*}{2}  & \multirow{3}{*}{4}  & bid                          & 69              & 115            & 66\%  \\
                               &                      &                               &                           &                     &                     & withdraw                     & 24              & 47             & 101\% \\
                               &                      &                               &                           &                     &                     & auctionEnd                   & 54              & 56             & 4\%   \\
\hline
\multirow{3}{*}{ERC721}        & \multirow{3}{*}{447} & \multirow{3}{*}{9}            & \multirow{3}{*}{13}       & \multirow{3}{*}{10} & \multirow{3}{*}{11} & transferFrom                 & 59              & 42             & -28\% \\
                               &                      &                               &                           &                     &                     & approve                      & 49              & 75             & 53\%  \\
                               &                      &                               &                           &                     &                     & setApprovalForAll            & 27              & 27             & 2\%   \\
\hline
\multirow{3}{*}{ERC20}         & \multirow{3}{*}{383} & \multirow{3}{*}{6}            & \multirow{3}{*}{18}       & \multirow{3}{*}{5}  & \multirow{3}{*}{6}  & transfer                     & 52              & 55             & 6\%   \\
                               &                      &                               &                           &                     &                     & approve                      & 47              & 50             & 7\%   \\
                               &                      &                               &                           &                     &                     & transferFrom                 & 43              & 50             & 15\%  \\
\hline
                               &                           &                     &                     &                              &                 & & & median:        & 14\%  \\
\hline
\end{tabular}
\caption{Overhead of Solidity programs generated by \proto, compared to reference implementations.
Column $\#Rules$ shows the number of rules in the declarative smart contracts.}
\label{tab:overhead}
\vspace{-5pt}
\end{table*}

\section{Evaluation}\label{sec:eval}

We implement a prototype compiler~\cite{dsc} for \proto in Scala 
that generates Solidity programs with instrumentation for run-time verification.
We first evaluate the compiler by comparing its output, 
without instrumentation, with reference contract written in Solidity. 
Next, we evaluate the overhead of run-time verification on these
contracts and their properties.

\subsection{Overhead to Reference Implementations}

\vspace{5pt}
\noindent \textbf{Reference smart contracts.}
We collect five reference smart contract implementations from
public repositories and prior research.
Wallet is the example shown in Section~\ref{sec:example}.
CrowdSale is from prior research paper~\cite{permenev2020verx}.
Auction is from Solidity documentation~\cite{solidity-auction}.
ERC20 (fungible tokens) and ERC721 (non-fungible tokens) are two of 
the most popular kinds of smart contract deployed on Ethereum,
\footnote{According to \url{https://etherscan.io},
at the time of writing this paper, there are about 502,000 
ERC20 tokens and 50,000 ERC721 tokens on Ethereum.}
and we use the implementation from the OpenZepplin 
library~\cite{openzeppelin}.

\vspace{5pt}
\noindent \textbf{Declarative smart contract implementation.}
We implement declarative counter-parts for all reference
contracts with the same interfaces and functionalities
without instrumentation for run-time verification or provenance.
These contracts consist of 10 to 18 rules 
(column \textit{\#Rules} in Table~\ref{tab:overhead}).

\vspace{5pt}
Although \proto can specify all the high-level logic of the
these contracts, we note that the generated Solidity code
has the following difference from the reference implementations.
First, the reference CrowdSale contract is implemented as two 
separate contracts. As \proto does not yet support contract
composition, the compiler outputs a stand-alone smart contract
with all the functionalities.
For the ERC721 contract, there is a \lstinline{safeTransferFrom}
interface, which wraps the \lstinline{transferFrom} function
with a check: if the recipient is also a smart contract, 
it should implement the \lstinline{onERC721Received} interface.
The current implementation of \proto does not yet support
such checking procedure, which relies on calling the built-in
functions of Solidity, so this interface is omitted.

\vspace{5pt}
\noindent \textbf{Measurement metrics.}
We measure two metrics: (1) the size of EVM byte-code deployed
on the blockchain; and (2) the gas cost for each transaction.
EVM byte-code is generated by the Truffle~\cite{truffle} compiler.
To measure gas cost, we first deploy the smart contract 
on Truffle's local blockchain, and then populate the smart contract states 
by sending transactions from $N$ test accounts, which results in $N$
entries in the contract states. Then we call each transaction
interface again and record gas cost reported by Truffle.
We find that $N$ (10 to 1000) does not impact gas cost.
This is because all contracts use hash-maps to store contract states.
If the hash-collision rate is low, the number of instructions is constant
to the size of the hash-map, and thus the gas cost remains constant.
Therefore, we report the gas cost measured with $N=10$.

\vspace{5pt}
\noindent \textbf{Results.}
As shown in Table~\ref{tab:overhead}, the median gas overhead to reference 
implementation is 14\% across 16 transactions,
with 3 of them have even lower gas cost between $-28\%$ to $-12\%$.
In the extreme case, the \lstinline{withdraw} transaction
from \lstinline{SimpleAuction} shows 101\% gas overhead.

\new{
We identify two sources of extra gas cost: (1) long function invocation chain, 
and (2) inefficient use of data structures.
For example, in the Wallet example (Section 2), mint transaction 
updates the variable \lstinline{totalMint}, which further updates 
\lstinline{totalSupply}, thus adding extra cost than directly 
incrementing \lstinline{totalSupply} as done by the reference implementation.
For data structures, relational tables are directly maintained as arrays of tuples, 
with extra information like valid bits and timestamps. 
Such extra information takes up additional space than their counterparts 
in Solidity implementations.
Mitigating such overhead borrowing ideas in SQL execution plan optimization
would be an interesting direction for future research. 
}

\new{
\proto consumes less gas in some transactions.
In Wallet, the \proto contract has less read / write to the global 
memory. In Crowfunding, the reference contract invokes an external call to another contract, 
whereas \proto implements everything in one monolithic contract, 
thus eliminating the inter-contract transaction cost.
In ERC721, \proto has fewer condition checks because some conditions
are specified as rules and therefore automatically maintained by the contract.
}

In terms of byte-code size, \proto's compiler output is slightly
greater than the reference programs, with a 2KB
(\lstinline{SimpleAuction}) maximum increase.
Note that on \lstinline{CrowSale}, \proto's output is
smaller than the reference contract. This is because the reference
implements two separate contracts, while the program generated
by \proto compiler has all functions implemented in one contract.


\vspace{5pt}
\noindent \textbf{Contract features that are not yet supported.}
During the search of benchmarks, we find some contracts use 
features that are not yet supported by \proto.
For example, the voting contract from Solidity documentation~\cite{solidity-voting} 
checks voting loop in a recursive manner.
Although recursion can be naturally expressed in \proto language, 
the execution of recursion functions requires non-trivial
reasoning to ensure termination and gas efficiency, and is therefore
not yet supported by \proto.
In addition, certain functions that lie outside of relational logic,
including checking interfaces of another contract 
(e.g. \lstinline{safeTransferFrom} in ERC721), 
and cryptographic functions\cite{solidity-purchase}, are not yet supported, 
but they can be incorporated into \proto via user-defined
functions in the future.

\newcommand\pwidth{2cm}

\begin{table}[]
\begin{tabular}{p{1.2cm}p{\pwidth}lll}
\hline
Contract      & Property                                & Size        & Transaction & Gas \\
\hline
\multirow{3}{*}{Wallet}        & \multirow{3}{\pwidth}{No negative balance}                     & \multirow{3}{*}{2} & mint                         & 14\%      \\
                               &                                                          &                    & burn                         & 14\%      \\
                               &                                                          &                    & transfer                     & 17\%      \\
\hline
\multirow{4}{*}{Crowdsale}     & \multirow{4}{\pwidth}{No missing funds}                        & \multirow{4}{*}{2} & invest                       & 50\%      \\
                               &                                                          &                    & close                        & 24\%      \\
                               &                                                          &                    & withdraw                     & 22\%      \\
                               &                                                          &                    & claimRefund                  & 33\%      \\
\hline
\multirow{3}{1cm}{Simple Auction} & \multirow{3}{\pwidth}{Refund once}                             & \multirow{3}{*}{2} & bid                          & 2\%       \\
                               &                                                          &                    & withdraw                     & 60\%      \\
                               &                                                          &                    & auctionEnd                   & 4\%       \\
\hline
\multirow{3}{*}{ERC721}        & \multirow{3}{\pwidth}{Every token has owner}                   & \multirow{3}{*}{1} & transferFrom                 & 5\%       \\
                               &                                                          &                    & approve                      & 3\%       \\
                               &                                                          &                    & setApprovalForAll            & 8\%       \\
\hline
\multirow{3}{*}{ERC20}         & \multirow{3}{\pwidth}{Account balances add up to total supply} & \multirow{3}{*}{1} & transfer                     & 96\%      \\
                               &                                                          &                    & approve                      & 13\%      \\
                               &                                                          &                    & transferFrom                 & 109\%     \\
\hline
                               &                                                          &                    & median:                      & 16\%     \\
\hline
\end{tabular}
\caption{Run-time verification overhead. Column $Size$ and $Gas$ 
show the overhead in byte-code size (KB) and gas cost (K) respectively, 
compared to the \proto contract without instrumentation.}
\label{tab:instrument}
\vspace{-10pt}
\end{table}

\subsection{Run-time Verification Overhead}
We measure run-time verification overhead by first specifying 
properties for each contract, which are generated as
instrumentation in the output Solidity program.
These instrumented programs are then
compared to \proto programs without instrumentation,
on byte-code size and gas usage.

Contract properties are specified as follows.
First, as shown in the example in Section~\ref{sec:example},
the Wallet contract is monitored for negative account balances.
The Crowdsale contract allows participants to invest in a 
crowd funding project with a particular funding target. 
The property specifies that the total amount of raised fund
should equal to all participants' investments.
In simple auction, bidders transfer their fund on every bid,
and get refunds when the auction is ended. A property specifies
that every bidder can claim refund at most once.
In ERC721, the property specifies that all
existing tokens should have a valid owner (non-zero address).
In ERC20, all account balances should add up to the 
total supply of tokens.

\noindent \textbf{Results.}
Table~\ref{tab:instrument} shows the overhead of run-time 
verification. Byte-code sizes are increased by no more than 2 KB.
Gas usage overhead varies across different transactions,
with the median being 16\%.
Wallet and ERC721 contracts show small overhead, 
where transaction gas consumption increases by no more than 17\% and 8\%, respectively.
Crowdsale and Simple Auction contract come with larger overhead.
The highest increase in their transaction gas usage are 50\% and 60\%.
The ERC20 contract shows the biggest overhead, where 
the \lstinline{transferFrom} transaction shows 109\% increase.

\section{Related Work} \label{sec:related}

In this section, we survey several lines of research that are related to our work.

\vspace{5pt}
\noindent \textbf{Run-time verification.}
Similar to \proto, Solythesis~\cite{li2020securing} also specifies properties as invariants
and generates instrumentation for run-time monitoring. 
It applies to general smart contracts implemented in Solidity, whereas \proto 
targets declarative contracts only.
By restricting the scope on declarative contracts, 
both specification and monitoring can be performed in a more straightforward manner. 
Invariants become violation queries, where joins are analogous to existential quantifiers, 
and aggregations to universal quantifiers. Detection becomes query evaluation,
which reuses the same procedure for contract execution.

SODA~\cite{chen2020soda} is a framework for implementing generic attack detection algorithms.
Unlike \proto, where the monitoring procedure is automatically generated from specification, 
the detection algorithms in SODA are implemented manually.

Sereum~\cite{rodler2020sereum} monitors reentrancy attacks online via taint analysis.
Azzopardi et al.~\cite{azzopardi2018observing} monitors contract execution against legal 
contract logic. These two work targets specific vulnerabilities and properties on Solidity 
smart contracts, whereas \proto monitors user-specified properties on declarative contracts. 

\vspace{5pt}
\noindent \textbf{Static analysis and verification.}
Static analysis has been applied to detect generic vulnerabilities such as 
reentrancy attacks \cite{grossman2017online,liu2018reguard}, 
integer bugs \cite{torres2018osiris,so2020verismart}, 
trace vulnerability \cite{nikolic2018finding}, 
and event-ordering bugs \cite{kolluri2019exploiting}.
Securify~\cite{tsankov2018securify} translates the EVM byte-code into 
stratified Datalog, and checks vulnerability patterns using 
off-the-shelf Datalog solvers. 

Alt et al.~\cite{alt2018smt} translates Solidity program into SMT formulas
and use off-the-shelf SMT solver to verify contract properties.
Zeus~\cite{kalra2018zeus} leverages abstract interpretations and
symbolic model checking to verify correctness and fairness of smart
contracts.

Symbolic execution~\cite{luu2016making,tsankov2018securify,albert2018ethir,
krupp2018teether,chang2019scompile,mossberg2019manticore,albert2019safevm,
permenev2020verx}
is another popular technique for smart contract verification.
Oyente~\cite{luu2016making} detects generic predefined vulnerabilities 
including reentrancy, transaction order dependency, mishandled exceptions, etc.
Verx~\cite{permenev2020verx}, on the other hand, allows programmers to specify 
contract-specific properties in temporal logic.

Fuzzing has also been applied to smart contracts.
For example, ContractFuzzer~\cite{jiang2018contractfuzzer} tests smart contracts
for security vulnerabilities.
Echidna~\cite{grieco2020echidna} generates tests that triggers
assertion violations.
ILF~\cite{he2019learning} and Harvey~\cite{wustholz2020harvey} 
focus on improving code coverage.

Unlike these work, \proto monitors properties online, 
which incurs run-time overhead,
but does not suffer from false-positives or false-negatives.
In addition, \proto targets declarative smart contracts, 
while these tools analyze Solidity or EVM byte-code.
Although targeting different languages,
the underlying verification techniques can also be applied to \proto and 
benefit from its higher-level abstraction.
We believe this is an exciting direction for future research. 

%
%
%
%

\vspace{5pt}
\noindent \textbf{Domain-specific languages for financial contracts.}
Scilla~\cite{sergey2019safer} is a intermediate-level language for smart contracts
that offers type safety and support for verification.
KEVM~\cite{Hildenbrandt2018kevm} defines the formal semantics of EVM, 
and has been used to verify contracts against ERC20 standards.
In contrast, \proto focuses on the high level abstraction of
smart contracts and specification of contract-specific properties.
\new{
Jones et al.~\cite{jones2003write} uses functional programming language to write
financial contracts.
}
BitML~\cite{bartoletti2018bitml} is a high-level language
for Bitcoin smart contracts.
Based on process calculus, it translates contracts into Bitcoin transactions.
\proto, on the other hand, is based on relational logic and targets Ethereum smart contracts.

\vspace{5pt}
\new{
\noindent \textbf{Datalog languages.}
\proto shares similar syntax with general Datalog languages
like Souffle~\cite{jordan2016souffle}, and is inspired by incremental evaluation techniques 
in systems like DDlog~\cite{ryzhyk2019differential}.
\proto however is specific to Ethereum smart contracts in the following aspects.
First, \proto has a number of domain-specific language extensions necessary for 
capturing execution semantics in Smart Contracts (Section~\ref{sec:language}).
Second, \proto compiles Datalog to Solidity, with several domain-specific optimizations 
(Section~\ref{sec:optimizations}).
Finally, \proto offers a property specification and run-time monitoring feature 
(Section~\ref{sec:instrument}), 
which is essential since smart contracts are managing a lot of digital assets.
}

\vspace{5pt}
\new{
\noindent \textbf{Deontic logic for normative knowledge.}
Gabbay et al.~\cite{gabbay2013handbook} presents a historical overview of
deontic logic for normative knowledge.
Based on similar principles, 
Prakken et al.~\cite{prakken2015law} overviews logic-based approaches 
for legal applications.
 \proto is a logical system representing knowledge 
in the domain of smart contracts, which enables efficient 
communication and automatic reasoning.
}

\section{Conclusion and future work}\label{sec:conc}

We present \proto, a declarative programming language for smart contract
implementation and property specification.
In \proto, smart contracts are specified in a high-level 
and executable manner, thus providing opportunities for efficient analysis 
and verification, bringing clarity to transaction execution via data provenance.
Contracts implemented in \proto demonstrate comparable efficiency to open-source
reference implementation.
Furthermore, run-time verification adds moderate gas overhead.

Our initial experience with \proto suggests a few exciting future directions.
First, we find interesting contracts that require additional language 
features, including contract composition, recursion, user-defined functions, etc.
Second, there are extreme cases where \proto compiler generates contracts 
with non-negligible overhead to the reference hand-written code.
\proto compiler needs further optimization to generate more efficient 
executable code.
Third, to save the overhead of run-time verification, 
we can leverage the high-level abstraction of \proto programs
to perform static verification.

\balance

\bibliographystyle{plain}
\bibliography{main,_blockchain,_system}

\end{document}